\documentclass[12pt]{article}

\usepackage{newtxtext,newtxmath}

\usepackage{graphicx}

\usepackage[letterpaper,margin=1in]{geometry}

\renewenvironment{abstract}
	{\quotation}
	{\endquotation}

\date{}

\makeatletter
\renewcommand{\fnum@figure}{\textbf{Figure \thefigure}}
\renewcommand{\fnum@table}{\textbf{Table \thetable}}
\makeatother

\usepackage[numbers,super,sort&compress]{natbib}

\usepackage{url}

\usepackage[font=small]{caption}
\title{A Joint Diffusion Approach to Multi-Modal Inference in Inertial Confinement Fusion}

\author{
Michael Jones$^{1}$\thanks{Corresponding author: jones313@llnl.gov} \and
Justin Kunimune$^{2}$ \and
Daniel Casey$^{1}$ \and
Bogdan Kustowski$^{1}$ \and
Eugene Kur$^{1}$ \and
Kelli Humbird$^{1}$
}

\date{
$^{1}$Lawrence Livermore National Laboratory, Livermore, CA 94550, USA\\
$^{2}$Plasma Science and Fusion Center, Massachusetts Institute of Technology, Cambridge, MA 02139, USA
}

\begin{document} 

\maketitle

\begin{abstract} \bfseries \boldmath
A combination of physics-based simulation and experiments has been critical to achieving ignition in inertial confinement fusion (ICF). Simulation and experiment both produce a mixture of scalar and images outputs, however only a subset of simulated data are available experimentally. We introduce a generative framework, called JointDiff, which enables predictions of conditional simulation input and output distributions from partial, multi-modal observations. The model leverages joint diffusion to unify forward surrogate modeling, inverse inference, and output imputation into one architecture. We train our model on a large ensemble of three-dimensional Multi-Rocket Piston simulations and demonstrate high accuracy, statistical robustness, and transferability to experiments performed at the National Ignition Facility (NIF). This work establishes JointDiff as a flexible generative surrogate for multi-modal scientific tasks, with implications for understanding diagnostic constraints, aligning simulation to experiment, and accelerating ICF design. 
\end{abstract}

\section*{Introduction}

Inertial confinement fusion (ICF) has emerged as a leading approach for controlled nuclear fusion~\cite{zylstra2022burning, abu2024achievement}, providing fundamental plasma physics insights and a potential pathway toward alternative energy. Scientific progress in ICF relies on a close interplay between simulation and experiment, in which each is iteratively refined based on data from the other. Machine learning models can enhance this interplay by acting as fast surrogates for simulations and by identifying input conditions that reproduce observed experimental outputs~\cite{anirudh2020surrogates, kustowski2022suppressing, gaffney2024data, spears2025predicting}. In practice, however, each implosion experiment yields a distinct and limited set of diagnostics, whereas simulations produce a richer and more uniform set of outputs. Moreover, these outputs are multi-modal -- a combination of scalars and images -- making it challenging to learn flexible conditional distributions based only on the available subset of diagnostics~\cite{kunimune20243d}. 

Multi-modal generative models have become powerful tools across scientific domains, enabling advances in biomolecular structure prediction, medical diagnostics, and materials design~\cite{abramson2024accurate, zhang2023biomedclip, wu2025versatile}. These models capture complex relationships across heterogeneous data types and often employ generative decoders such as denoising diffusion probabilistic models (DDPMs)~\cite{ho2020denoising, song2020denoising} to generate high-quality samples. Multi-modal prediction is typically achieved either by combining data into a shared latent space prior to diffusion~\cite{bao2023one, li2025omniflow} or by aligning pre-existing latent representations using contrastive learning techniques~\cite{radford2021learning}. While contrastive approaches allow the reuse of pre-trained models without exhaustive retraining, they may fail to capture subtle inter-modal dependencies by not explicitly modeling the full joint distribution~\cite{chen2024multi}.

An alternative strategy is to train a generative model directly on the joint distribution of multi-modal data. This approach has gained traction in domains such as generative biology, where diffusion models jointly update continuous atom positions and discrete atom types~\cite{chu2024all, qu2024p, campbell2024generative}. It has further been shown that joint diffusion is feasible and theoretically well-founded in arbitrary state spaces \cite{holderrieth2024generator, rojas2025diffuse}, however the application of joint diffusion models as physics-based surrogates have been under-explored.

We propose JointDiff, a model architecture and training scheme that leverages joint diffusion as a generative surrogate model for ICF. We train our model on a large ensemble of 3D Multi-Rocket Piston (RP) simulations~\cite{casey2025multi} and demonstrate accuracy in forward and inverse modeling tasks as well as imputation of missing outputs. We show that JointDiff produces meaningful conditional distributions from partial multi-modal observations, enabling tunable uncertainty quantification and investigation of input sensitivity to missing diagnostics, with demonstrated transferability to experiments at the National Ignition Facility (NIF).

\begin{figure}
\centering
\includegraphics[width=0.8\linewidth]{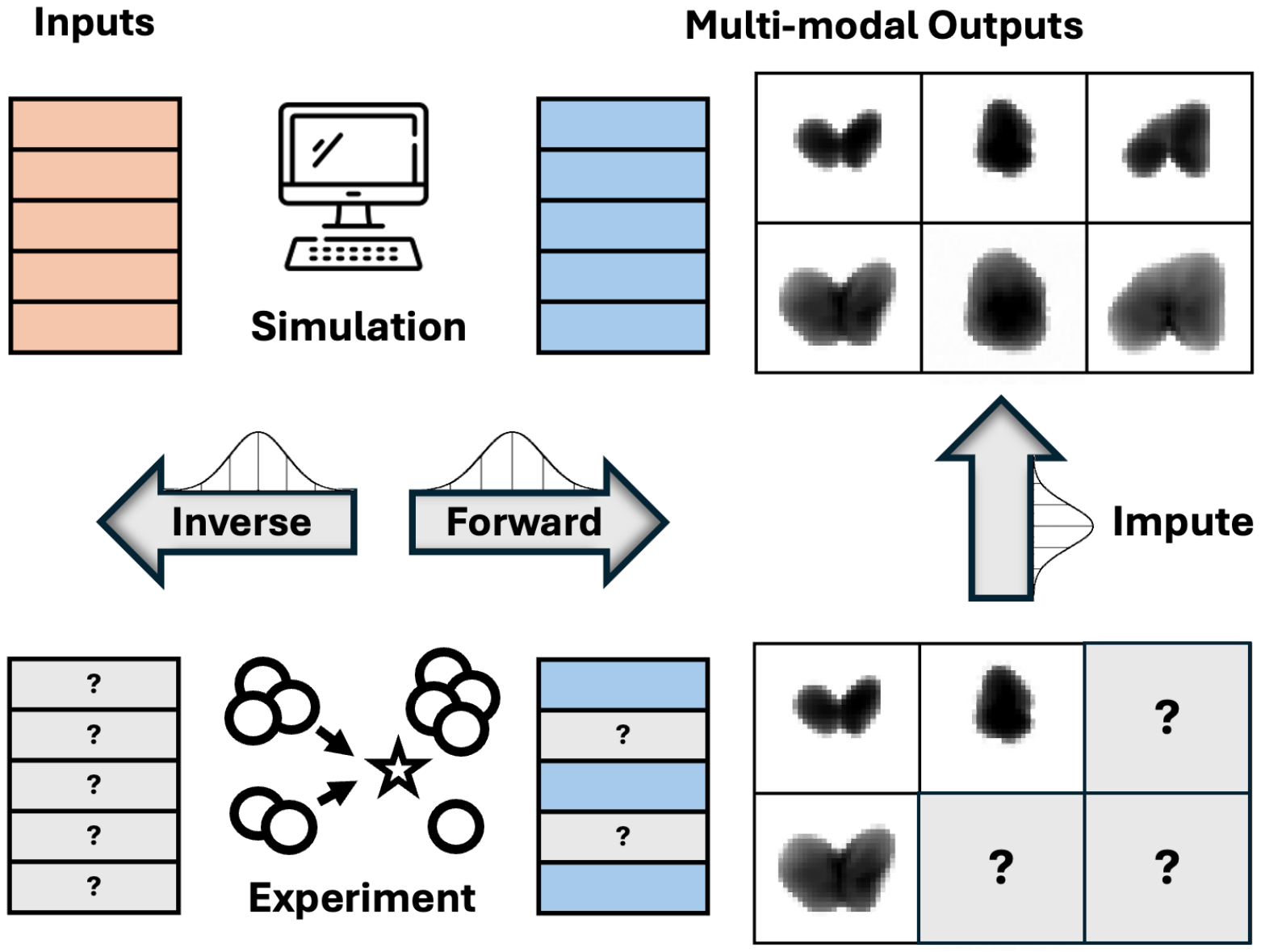}
\caption{\textbf{JointDiff is generative surrogate for forward, inverse, and imputation tasks.} Simulations (upper row) produce a complete set of multi-modal outputs given a complete set of inputs. Surrogate models are trained to replace simulations and perform the forward or inverse predictions tasks. Experiments (bottom row) produce incomplete sets of outputs, and inputs to the capsule are unknown. In addition to forward and inverse surrogate capabilities, JointDiff predicts conditional distribution of inputs and outputs given partial multi-modal observations.}
\label{fig:app}
\end{figure}

\begin{figure}
\centering
\includegraphics[width=0.8\linewidth]{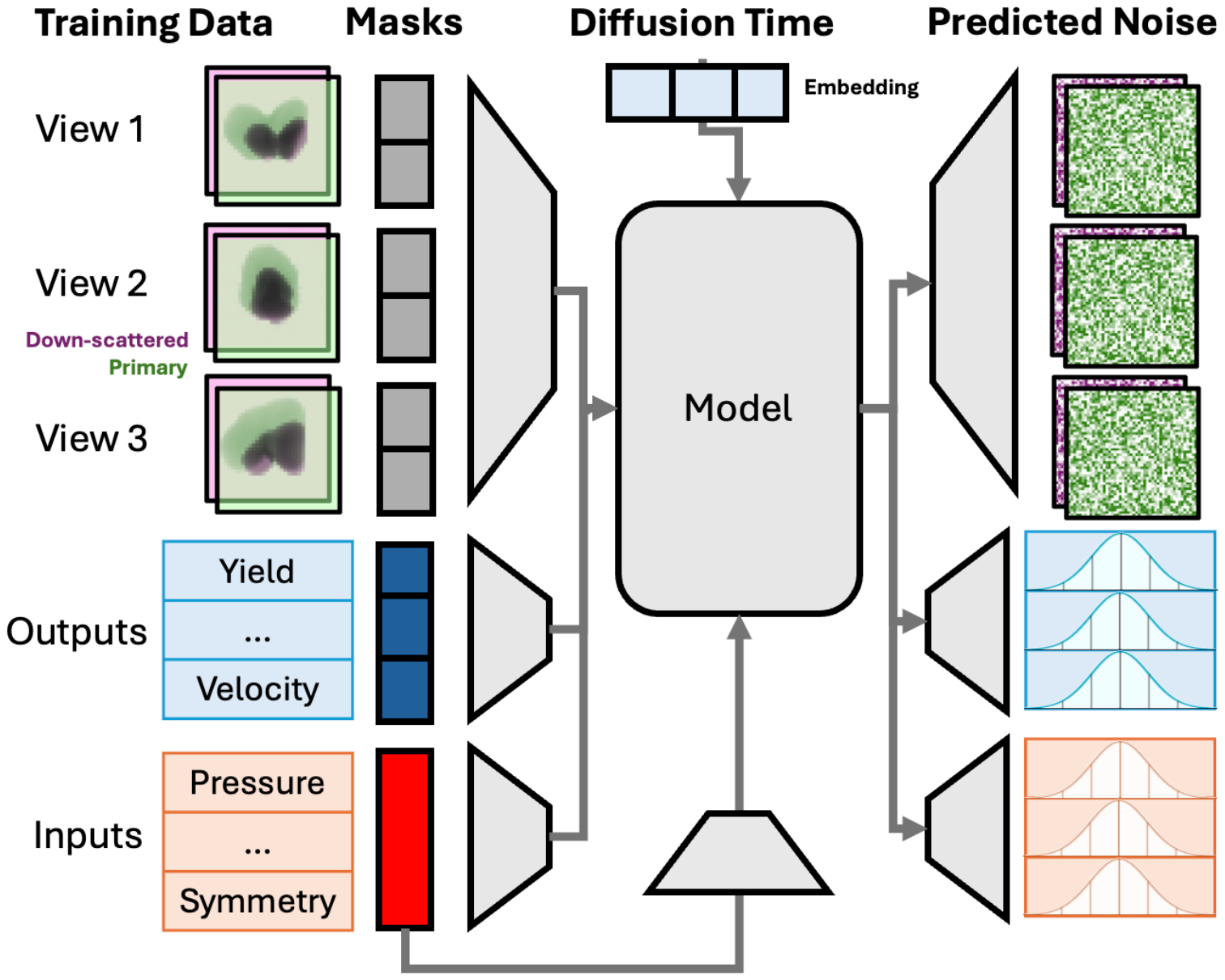}
\caption{\textbf{Architecture of the JointDiff model.} Outputs images contain primary (green) and down-scattered (pink) neutron intensities, which are encoded as separate image channels. Trapezoids represents data encoders and decoders, which are convolutional networks for images and fully connected networks for scalars data. Gray and blue boxes correspond to discrete masks which inform the model if true or noisy data is provided for each modality and are embedded by a fully connected network. Diffusion time is encoded by a sin/cos positional embedding.  The time, mask, and input/output embeddings are concatenated to the image embeddings at each convolution step. Individual decoders predict noise to be removed at diffusion time $t$ for each image and set of input and output scalars.}
\label{fig:arch}
\end{figure}

Deep learning models have previously been applied as multi-modal ICF surrogates~\cite{anirudh2020surrogates, kustowski2022suppressing} and techniques such as Markov Chain Monte Marlo (MCMC) have been used to reconstruct simulations inputs from experimental outputs~\cite{gaffney2024data, kunimune20243d, spears2025predicting}, however implementing MCMC with multi-modal data is sensitive to choice of priors, computationally intensive, and may fail to converge. DDPMs have been applied as surrogates for Particle-in-cell (PIC) simulations~\cite{liu2024diff}, which are crucial to understand fundamental plasma interactions during ICF implosions, and preliminary work showed that DDPMs can model 2D radiation hydrodynamics simulations outputs with single-channel images~\cite{jonestowards}. In this work, we generalize this framework to 3D simulation data using an multi-modal U-net based architecture amenable to arbitrary image views, images channels, and scalars inputs and outputs (Figure \ref{fig:arch}). 

We evaluate JointDiff on three core prediction tasks: (1) surrogate modeling (simulation inputs $\rightarrow$ outputs), (2) inverse modeling (outputs $\rightarrow$ inputs), and (3) imputation of missing outputs (partial outputs $\rightarrow$ complete outputs and inputs). Each task (shown in Figure \ref{fig:app}) addresses a critical component of the ICF design process -- accelerating simulations for new input condition, inferring inputs for previous experiments, and enriching available outputs -- leading to the development of higher-yield and more robust experiments. We perform round-trip consistency tests to demonstrate that the learned distributions remain stable and self-consistent even when a large fraction of simulation data is masked. Finally, we validate our approach using recent ICF experiments performed at the NIF, which contain only partial scalar and image diagnostics relative to simulations. Without fine-tuning on experimental data, JointDiff’s round-trip predictions closely reconstruct most experimental observables. We also note discrepancies in specific scalar and image features which provide insight into limitations of the underlying RP physics model. Together, these results show that JointDiff is a robust and flexible framework for predicting conditional distributions in ICF, and, more broadly, in scientific domains characterized by partial multi-modal observations.

\section*{Results}

\subsection*{Joint diffusion enables multi-modal conditional prediction}


To predict conditional distributions given multi-modal (scalar and image) data, we use a joint diffusion objective \cite{holderrieth2024generator, radford2021learning}. During training, noise is gradually added to both scalars and images and a single architecture learns to ``denoise'' both modalities at the same time. We apply random masking over inputs and outputs such that the models learns the distribution of outputs given inputs (forward model) as well of inputs given outputs (inverse model). Output masks are varied individually over scalars and images, enabling prediction of inputs and missing outputs conditioned on partial observations. When making predictions, the model is guided by whatever data is available, and the prediction targets are masked. The model architecture is shown in Figure \ref{fig:arch} and additional details are provided in Methods. 

We train our model on over 443 thousand simulations generated by a RP physics model. The RP model uses a discretized shell to describe how fuel in a target capsule is compressed to form a hotspot;  more detail can be found in Methods and in~\cite{casey2025multi, kunimune20243d}. Each simulation training example contains a set of 28 scalar inputs, 12 output scalars, and 3 image-line-of-sight each with 2 energy channels. The input scalars represent parameters such as initial pressure, adiabat, and drive symmetry modes that define each simulation. Scalar outputs include quantities that can be directly compared to experiment, such as the total neutrons generated by the implosion (Yield) and the time of peak neutron production (Bang-time), as well as inferred quantities that cannot be measured experimentally such as the areal density ($\rho$R) and residual kinetic energy (RKE), which are measures of hot spot confinement and energy not coupled to the hot spot, respectively. The images correspond to three different views of the implosion, with each image containing two channels: one for higher-energy (Primary) neutrons and one for lower-energy (Down-scattered) neutrons. A complete list of inputs and outputs can be found in the Supplementary Text.

\subsection*{JointDiff is an accurate surrogate for Rocket-Piston Simulations}

\begin{figure}
\centering
\includegraphics[width=0.8\linewidth]{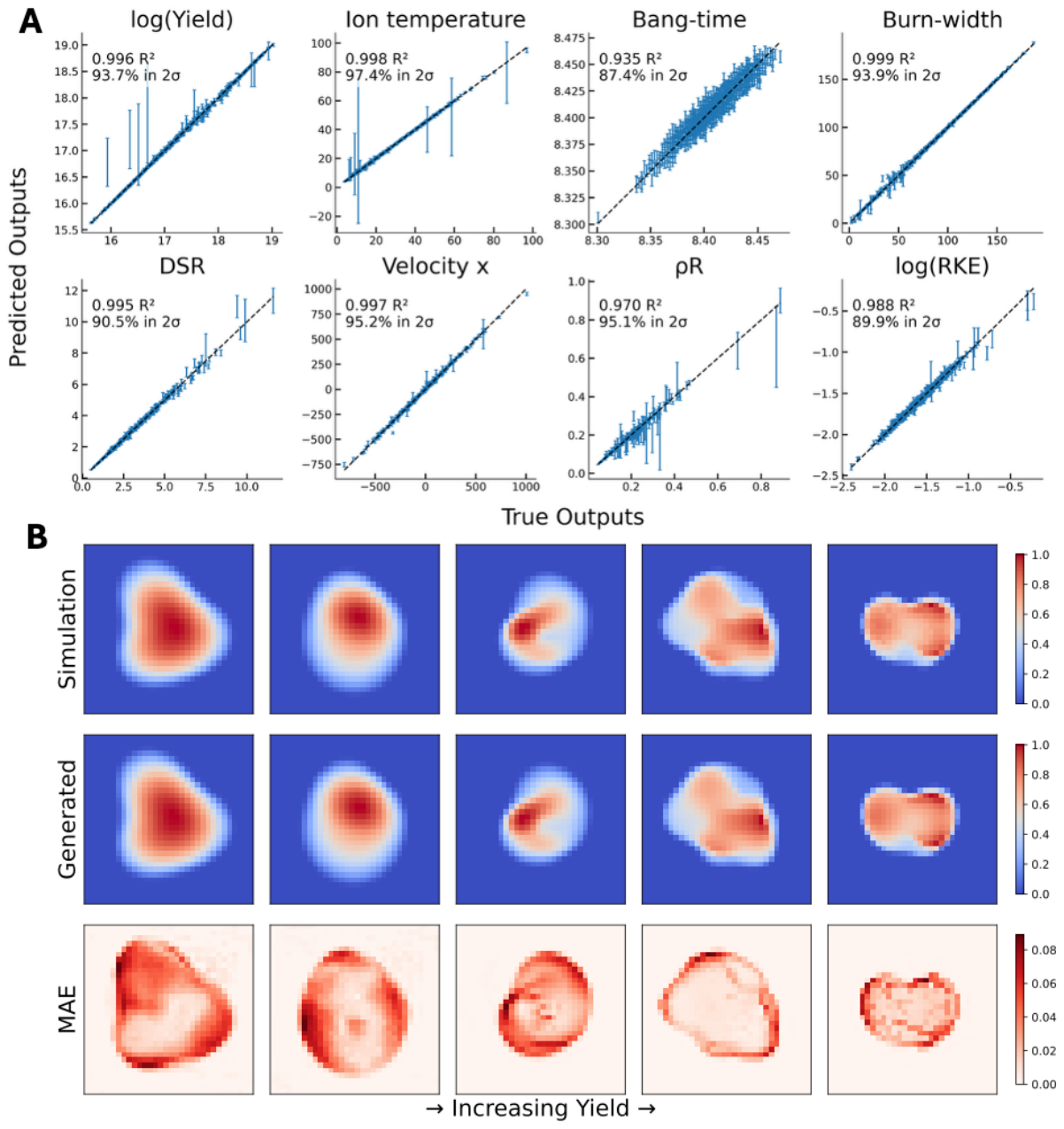}
\caption{\textbf{JointDiff predicts accurate distributions of RP simulation outputs.} A) For each set of test inputs, 10 outputs predictions are made. Error bars show the mean and standard deviation across these 10 predictions. $R$2 metrics are computed between the predicted means and the simulated ground truth. The percent of ground truth samples within two standard deviations of the predicted distribution is also shown for each scalar. A subset of 8 outputs are shown with all inputs in Figure \ref{fig:outputs_all}. B) Primary neutron images (View 1) sampled from each quintile of yield in the test data. The first row shows the ground truth simulated images, the second row shows the mean model prediction across generated samples, and the third row shows MAE between each generated sample and the ground truth.}
\label{fig:fwd_inv}
\end{figure}

We first evaluate JointDiff by predicting simulation outputs given scalar inputs for 986 test samples excluded from training. In Figure \ref{fig:fwd_inv}A we show mean and standard deviations given 10 output predictions for each test sample (additional velocities and inferred outputs shown \ref{fig:outputs_all}). JointDiff achieves excellent prediction accuracy, with $R^2$ values ranging from 0.935 to 0.999 across scalars. On average, 92.8\% of true samples fall within two standard deviation of the predicted distribution, varying 84.7\% to 97.4\% across scalars. This is close to the 95.5\% expected for a Gaussian distribution, however we emphasize that predicted distribution are not Gaussian and calibration varies by scalar. Still, we observe a significant majority of ground truth outputs fall within predicted distribution. Additionally, JointDiff can help identify outlier samples. For instance, the four samples with the largest yield prediction errors also exhibit the greatest yield uncertainty. This demonstrates how predicted distributions can flag less trustworthy predictions, in contrast to deterministic models that cannot directly quantify uncertainty.

We next evaluate the model's ability to generate accurate neutron images conditioned on simulation inputs. In Figure \ref{fig:fwd_inv}B we compare ground truth primary neutron images (View 1) to the mean of 10 images generated for five test set samples. Each test image is randomly sampled from a different quintile of total neutron yield, which spans three orders of magnitude and results in distinct intensity profiles. Despite the diversity in yield and 3D geometry, the model consistently generates detailed and visually similar images across the sampled range. When analyzing the mean absolute error (MAE) for each generated image compared to the ground truth, we observe that the largest pixel-wise errors occur at the boundaries of the intensity profiles. Since these inter-generation discrepancies are not apparent in the mean images, they likely correspond to regions of higher variance which can be used to identify regions of lower confidence in the predicted images.

\subsection*{JointDiff predicts inputs and missing outputs given multi-modal conditioning}

We test the capabilities of JointDiff when guided by multi-modal output scalars and image information, focusing on two tasks: imputing missing scalars and predicting input distributions. Imputation is particularly relevant to scientific applications where simulation outputs may be unobservable in experiments or intermittently unavailable due to measurement challenges. This is a common occurrence in ICF experiments, where particular diagnostics might fail, are blocked by ride-along experiments, or are physically incapable of resolving simulation outputs. In Table \ref{tab:impute_outputs} we evaluate predicted distributions for five outputs that were left out (masked) during inference. Similar to the outputs analysis above, we obtain standard deviations over 10 diffusion model generations for each test sample, and we observe similar or slightly improved accuracy as well as similar calibration to forward model predictions shown in Figure \ref{fig:fwd_inv}. Improved performance compared to the forward model is likely due to stronger correlation between outputs that serve to constrain the missing values. We note that Bang-time has the lowest accuracy and highest uncertainty in both cases, indicating more noise in the simulated diagnostic or lower correlation with other outputs.

For the same test set, we mask inputs and guide conditioning with all output scalars and images. In Table \ref{tab:predict_inputs} we show predicted distributions for five input scalars (all scalars shown in \ref{fig:inputs_all}) compared to their ground truth values. Again we observe strong accuracy across scalars, with $R^2$ ranging from 0.967 to 0.999, and find that a majority of predicted standard deviations include the ground truth scalar with a mean of 93.7\% within two standard deviations. For both the inverse and forward modeling tasks we perform ablations with task-specific and deterministic variants of JointDiff and show comparable performance (Table \ref{tab:baselines}).

\begin{figure}
\centering
\includegraphics[width=0.6\linewidth]{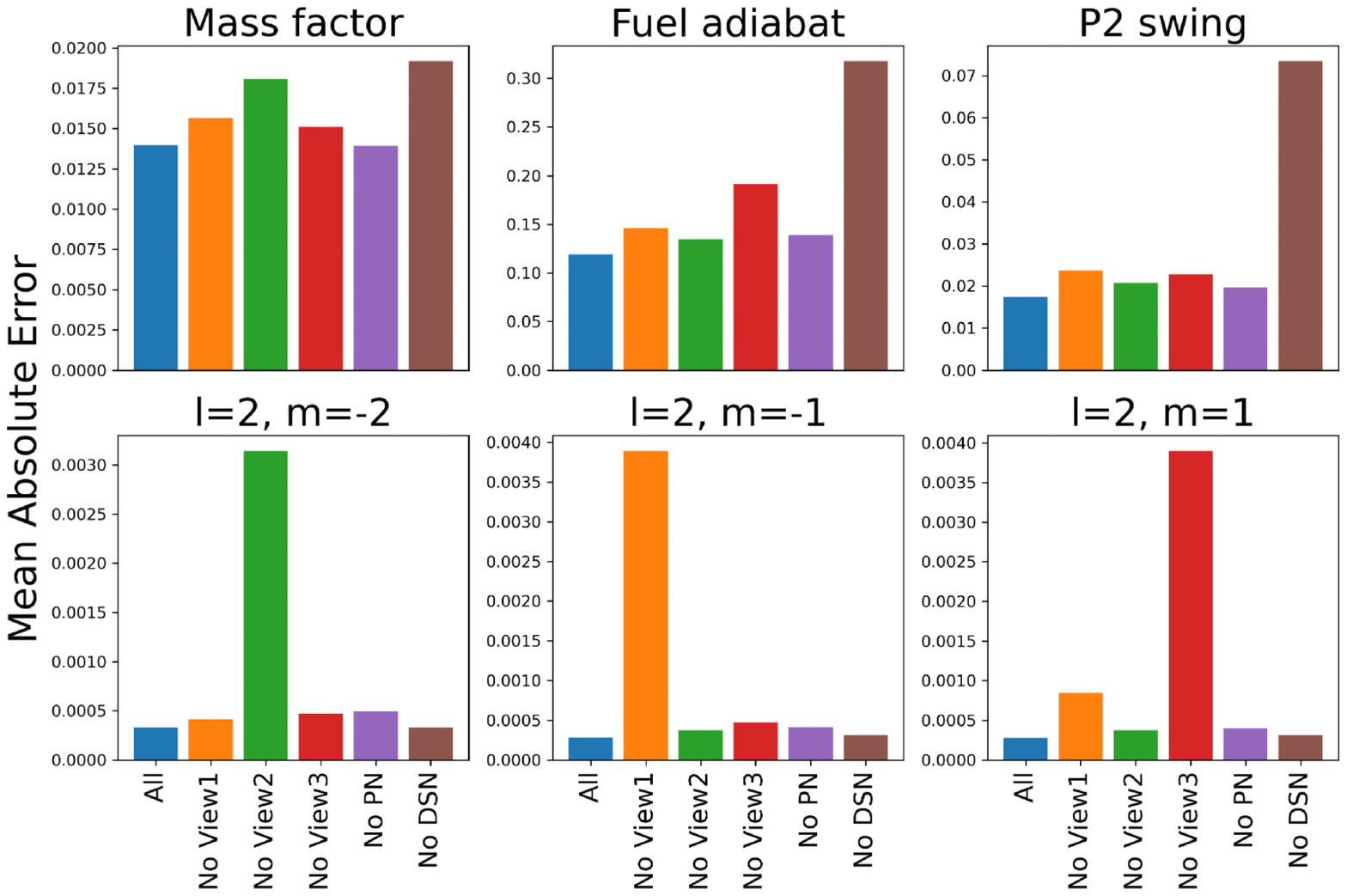}
\caption{\textbf{Sensitivity of input predictions to removal of outputs image views and channels.} Mean average error for six inputs when providing all outputs (blue) and when removing partial image information (each view and each channel). All inputs are shown in the Figure \ref{fig:mae_all}}
\label{fig:mae}
\end{figure}

In Figure \ref{fig:mae} we test the inverse model's robustness and sensitivity removing partial image information. For six inputs (all inputs shown in \ref{fig:fig5_inps_combined}) we show the prediction MAE when removing each individual image view, as well as all primary or all down-scattered neutron images. In most cases, there is minimal change to the MAE when partial information is removed, however we find that particular inputs are sensitive to the removal of particular images. For example, removing View 3 increases the l=2,m=1 symmetry error by 10x, while having minimal impact on l=2,m=-1. Removing View 1 has a similar impact on l=2,m=-1 while preserving the accuracy of l=2,m=1. These inputs represent diagonal drive symmetry features, and this analysis reveals that they cannot both be resolved without access to both equatorial views (1 and 3). Furthermore, we find that excluding down-scattered images has the greatest effect on adiabat and P2 swing predictions. This is likely because down-scattered images offer the most insight into the confining shell’s profile -- the adiabat sets the shell's density, and the P2 swing strongly changes the shell's density distribution. These correlations are important to consider when  1) gauging uncertainty for experiments with missing diagnostics and 2) designing new diagnostic capabilities at the NIF and future facilities.

\subsection*{Conditional inputs distributions remain self-consistent in the absence of multiple outputs}

Next we test the inverse model's ability to adapt and predict inputs when a larger fraction of output data is unavailable. In this test we specifically remove outputs that are regularly unavailable in NIF shots. For example, in NIF experiments View 3 is often blocked by other diagnostics and the View 2 (polar) detector cannot capture down-scattered neutrons as the camera's line of sight is not long enough to separate neutron energies in time~\cite{fatherley2017system}. Accordingly, we mask both channels of View 3 as well as the down-scatter channel of View 2. Furthermore, 4/12 scalars in our simulation dataset, measuring $\rho$R and RKE quantities, are inferred outputs that have no direct experimental equivalent. Burn-width measurements can also be challenging at high yield where burn duration is comparable to detector resolution, and the RP Bang-time tends to be systematically lower than experiments~\cite{casey2025multi}. Therefore we remove half (6/12) of scalar outputs and half (3/6) of image channels, and we test the model's ability to predict meaningful capsule inputs distribution.  We emphasize that this model was not re-trained or fine-tuned for this specific mask configuration and it therefore retains the forward and inverse modeling capabilities showcased above.

\begin{figure}
\centering
\includegraphics[width=0.8\linewidth]{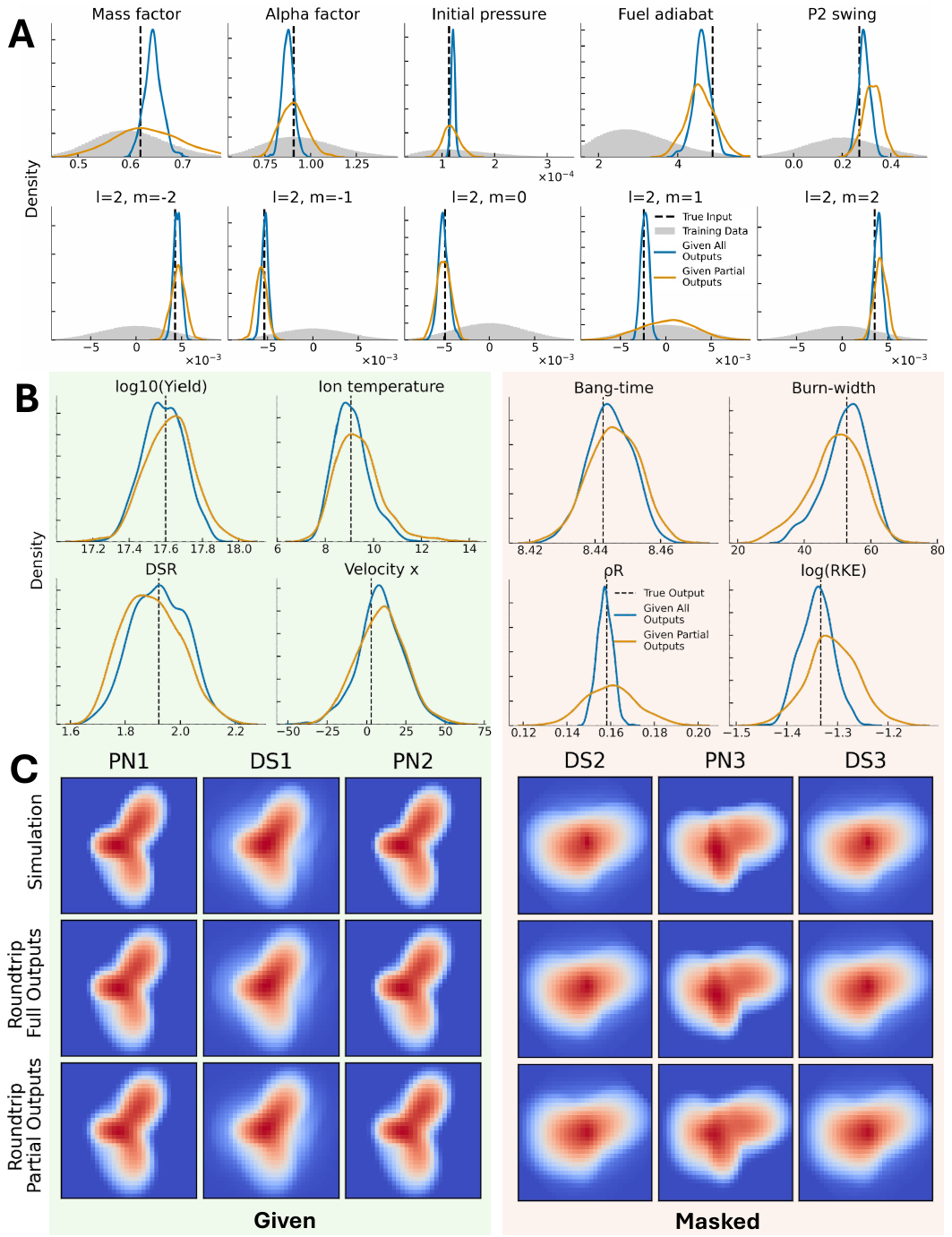}
\caption{\textbf{Input and round-trip distributions for single test sample.} A) Distribution of inputs for 500 generations when providing model all outputs (blue) or half of outputs (orange). Ground truth inputs are shown by black dashed line. Distribution of all training data is shown in gray. B) Round-trip output distributions produced by running inputs from A through the forward model. Both round-trip distributions of given outputs (green shading) are similar and centered on the original outputs, despite originating from different inputs distribution. Round-trip distributions for masked outputs (orange shading) are less consistent but retain significant overlap. C) Round-trip images produced by running inputs from A through the forward model. Simulated images (first row) are compared to mean images of full outputs (second row) and partial outputs (third row).}
\label{fig:rt_sim}
\end{figure}

In Figure \ref{fig:rt_sim}A we show a distribution of inputs given 500 generations all conditioned on a single set of simulation outputs. We highlight all l=2 symmetry modes as these are main contributors performance degradation at NIF. We compare distributions generated from full knowledge of scalar and image outputs (blue) to those generated from partial knowledge (orange). For inputs such as the P2 swing and most symmetry modes, these distributions look very similar, indicating that adding the missing scalars and images views does not enhance confidence in input predictions. Other quantities, such as the initial pressure and alpha factor, have wider distributions with slightly shifted means when partial information is given, indicating that adding the missing information is helpful in constraining these distributions beyond what is commonly observed in experiments. Interestingly, we note that the l=2,m=1 mode collapses entirely on to the training data distribution when given partial outputs, whereas it is tightly constrained around the ground truth input when given full outputs. This result makes sense in the context of Figure \ref{fig:mae} where we observe a much larger MAE for the l=2,m=1 mode when view 3 is removed. We show additional examples in the complete input space in SI4 which show similar trends to Figure \ref{fig:rt_sim} with some expected variation given differences in image profile and scalar magnitudes. This analysis provides guidance for both the relative utility of various ICF diagnostics and the confidence in input predictions when a given diagnostic fails.

Given the high dimensionality of the input space, it is challenging to verify that the predicted input distribution appropriately describe the outputs they are conditioned on. Indeed we observe that the ground truth inputs (vertical black lines) tend to fall within both the full and partial conditional distributions, but these represent only a single solution that might describe the observed output. In order to evaluate the complete distribution, we enlist the forward version of the model (which was shown to be a highly accurate surrogate in Figure \ref{fig:fwd_inv}) to project samples back into outputs space. This ``round-trip'' analysis, visualized in Figure \ref{fig:rt_sim}B (additional examples shown in SI5), shows that both the partial outputs and full output distributions are self-consistent with the original conditioning values, as demonstrated by distributions centered on the ground-truth black dashed lines. For the scalar outputs that are provided in both cases (green shading) the round-trip distributions are very similar in both mean and variance. For outputs that were missing in the partial distribution (orange shading) distributions are similar for Bang-time and Burn-width, but the partial distributions are wider for $\rho$R and RKE, indicating that the wider inferred distributions do not adequately constrain these values.

We perform the same round-trip analysis on images. Images channels provided to both models are shown in the green box in Figure \ref{fig:rt_sim}C, and masked images are shown in the red box. Again we observe strong round-trip reconstructions when comparing the mean round-trip prediction to the original images, although there are some features slightly blurred in the partial round-trips. Taken together, these results verify that the predicted inputs distributions correctly correspond to their conditioning values and that the forward and inverse modes are self-consistent, even when given partial information and without strict self-consistency enforced as in~\cite{anirudh2020surrogates}.

\subsection*{JointDiff is transferable to NIF Experiments}

\begin{figure}
\centering
\includegraphics[width=1.0\textwidth]{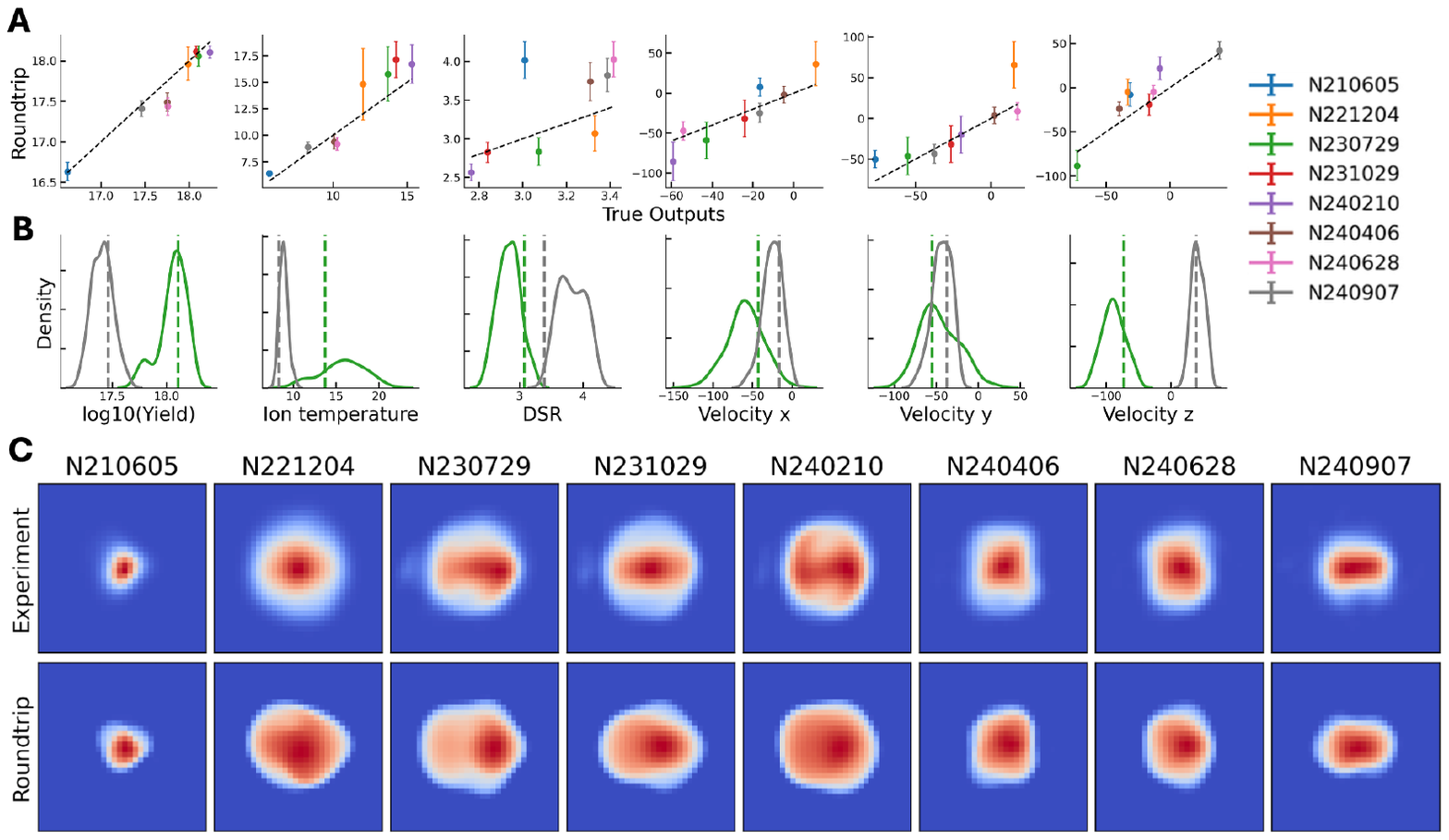}
\caption{\textbf{Round-trip predictions given partial experimental outputs.} A) Round-trip mean and standard deviations for six output scalars measured in eight NIF experiments. B) Round-trip output distributions over 100 generations for N230729 (green) and N240907 (gray) experiments. Experimental outputs values are shown as dashed lines in green and gray, respectively. C) Experimental (top row) and round-trip generated images (bottom row) of View 1 primary neutrons for all eight shots.}
\label{fig:exp}
\end{figure}

After verifying that the model satisfies round-trip tests for partial simulations obervables, we perform the same analysis on a set of NIF experiments. Each experimental shot contains data for the six scalars and three image channels described above, and can therefore be used to produce an analogous input and round-trip output distributions. We emphasize that no experimental data was used to train or fine-tune the JointDiff model, therefore we do not expect the model to reproduce various experimental features that are not captured by the relatively simple RP model. Indeed, if experimental data were significantly out-of-distribution compared to RP simulations and the JointDiff model were not robust, we would expect the round-trip distributions to be entirely uncorrelated with the observed experimental outputs they were conditioned on. However, in Figure \ref{fig:exp}A we observe reasonably strong correlations for most scalars, indicating the model is transferable to experiment.

In Figure \ref{fig:exp}B we examine the round-trip distribution of scalars and images predicted for the recent N230729 and N240907 shots. We observe that predicted outputs distributions are wider than simulations, indicating higher uncertainty, but that ground-truth outputs are captured within the distributions. Across all shots, DSR is the hardest scalar to match, potentially indicating inconsistency in the underlying RP model. In Figure \ref{fig:exp}C we show round-trip predictions of View 1 primary neutron images for all experiments. Overall image shape is well captured in reconstructions, however more subtle features are blurred or lost, especially in N221204 and N240210. These features may not be present in the RP training data, and fine-tuning JointDiff on experiments or higher fidelity simulations~\cite{humbird2021cognitive, kustowski2022suppressing} would likely improve round-trip self-consistency.

This analysis provides a means of assessing both JointDiff robustness to out-of-distribution data as well as the physical similarity of RP training data to experiments. Importantly, we show that the model can produce an input distribution that approximately reproduces observed outputs. These analysis may guide future development of the RP model -- for example to align DSR predictions -- and sampling strategies to approach experimental conditions near the ignition cliff.

\section*{Discussion}

We presented the JointDiff model for multi-modal conditional generation, demonstrating its ability to accurately model distributions in three key scenarios: generating multi-modal outputs from simulation inputs, inferring simulation inputs from multi-modal outputs, and imputing outputs from partial observations. We showed that predictions are statistically meaningful and and we demonstrate self-consistency even when 50\% of scalars and images are masked. Finally we show that the model and underlying RP training data supports predictions on fusion experiments performed at the NIF.

Looking ahead, several directions may further improve model performance and generalizability. For ICF applications, we plan to fine-tune JointDiff using higher-fidelity simulation data from codes such as HYDRA~\cite{marinak2001hydra}, as well as experimental data. For more complex datasets, future work could incorporate the joint diffusion objective into advanced architectures like the diffusion transformer~\cite{peebles2023scalable, rojas2025diffuse}, apply curriculum learning strategies~\cite{bengio2009curriculum} to incrementally introduce challenging mask configurations, or project images into pre-trained latent spaces before diffusion~\cite{rombach2022high}. Finally, integrating experimental uncertainties during inference may help constrain predicted distributions and better align model outputs with results from MCMC approaches.

Overall, our findings demonstrate that JointDiff provides a robust and versatile framework for ICF that may be generally applicable for multimodal conditional generation tasks across scientific domains.

\clearpage

\begin{table}
\centering
\caption{\textbf{Imputing missing scalar outputs given all other scalars and image outputs} Standard deviations are computed  across 10 generations for each sample in the test set.}
\label{tab:impute_outputs}
\begin{tabular}{lcc}
\\
\hline
Output & $R^{2}$ & Within $2\sigma$ \\
\hline
log(Yield) & 1.000 & 95.5 \\
Ion temp   & 1.000 & 97.9 \\
Bang-time  & 0.932 & 87.8 \\
Burn-width & 1.000 & 92.5 \\
DSR        & 0.999 & 95.1 \\
\hline
\end{tabular}
\end{table}

\begin{table}
\centering
\caption{\textbf{Predicting simulation inputs from scalar and image outputs (inverse model)} Drive symmetry is averaged across all 23 symmetry modes. Standard deviations are computed  across 10 generations for each sample in the test set.}
\label{tab:predict_inputs}
\begin{tabular}{lcc}
\\
\hline
Input & $R^{2}$ & Within $2\sigma$ \\
\hline
Mass factor  & 0.969 & 96.7 \\
Alpha factor & 0.984 & 96.2 \\
Fuel adiabat & 0.987 & 97.9 \\
P2 swing     & 0.987 & 98.7 \\
Symmetry    & 0.991 & 95.8 \\
\hline
\end{tabular}
\end{table}


\clearpage 

%
\bibliography{science_template} 
\bibliographystyle{unsrt}

%
%
%
%
%
%


\section*{Acknowledgments}
The authors thank the international Inertial Confinement Fusion collaboration and the LLNL ICF program.
\paragraph*{Funding:}
This work was performed under the auspices of the US Department of Energy by Lawrence Livermore National Laboratory under contract DE-AC52-07NA27344.
\paragraph*{Author contributions:}
K.H., E.K., B.K. and M.J. designed the research. M.J performed research. D.C. and J.K. generated simulation training data. M.J. wrote the paper. K.H., D.C, B.K., and J.K. advised the paper.
\paragraph*{Competing interests:}
There are no competing interests to declare.
\paragraph*{Data and materials availability:}
Relevant NIF data is included in the text, figures, and supplemental information of this manuscript. Simulation and model data maybe be available upon reasonable request to the authors.

\subsection*{Supplementary materials}
Materials and Methods\\
Figures S1 to S5\\
Tables S1 to S3\\


\newpage


\renewcommand{\thefigure}{S\arabic{figure}}
\renewcommand{\thetable}{S\arabic{table}}
\renewcommand{\theequation}{S\arabic{equation}}
\renewcommand{\thepage}{S\arabic{page}}
\setcounter{figure}{0}
\setcounter{table}{0}
\setcounter{equation}{0}
\setcounter{page}{1} 


\begin{center}
\section*{Supplementary Materials}


\end{center}

\subsubsection*{This PDF file includes:}
Materials and Methods\\
Figures S1 to S5\\
Tables S1 to S3\\

\newpage


\subsection*{Materials and Methods}

\subsubsection*{Multi-Rocket Piston model}
\label{sec:methods_RP}

Simulated implosions are modeled using a radiation drive derived from the Callahan hohlraum model~\cite{callahan2020simple} and solve the rocket equations for ~200 coupled rocket-pistons that discretize the capsule shell in 3D. After the initial acceleration phase, the pistons are linked through hotspot pressure using power-balance equations, following the framework of Springer et al.~\cite{springer20183d} but with an explicit hohlraum model applied from the onset of implosion. Synthetic diagnostics at stagnation and peak burn are generated through post-processing. Full methodological details are provided in~\cite{casey2025multi}.

\subsubsection*{Training Data}
An ensemble of 443,610 RP simulations was used with inputs sampled near the predicted experimental ignition cliff~\cite{casey2025multi}. A random split of 986 samples (0.2\%) was held out for testing, and 100 additional samples were used for validation during training. Images were normalized to $[0, 1]$ by dividing by max pixel intensity for each view and channel. Scalar inputs and outputs were normalized to $[0, 1]$ given mean and then passed through an inverse sigmoid $\hat{x} = \log{x}/(1-x)$ to more closely approximate zero-centered Gaussians. All evaluation metrics reported in Tables \ref{tab:impute_outputs}  and \ref{tab:predict_inputs} and Figure \ref{fig:fwd_inv} are performed after removing normalization.

\subsubsection*{ICF Experiments}
Deuterium-Tritium ICF experiments were performed at the National Ignition Facility (NIF) between June 2021 and September 2024. Additional information on specific shots can be found in~\cite{abu2022lawson, abu2024achievement, hurricane2025inertial, zylstra2022experimental, kritcher2024design, hurricane2024present}.

\subsubsection*{Denoising Diffusion Probabilistic Model}
DDPMs generate high resolution data distributions by first noising training data to gradually approach a Gaussian distribution, then iteratively denoising samples drawn from a Gaussian prior~\cite{ho2020denoising, song2020denoising}. During training, samples $X_t$ can be drawn at arbitrary diffusion times $t$ according to $X_t = \sqrt{\bar{\alpha_t}} X_T + \sqrt{1-\bar{\alpha_t}} \epsilon$, where $t \sim [0, 1]$, $\epsilon \sim \mathcal{N}(0, 1)$ and $\bar{\alpha}_t = \prod_{s=1}^{t}\alpha_t$ corresponding to the noise schedule. We use a linear noise schedule that decreases $\alpha_t$ from 0.9999 to 0.98 over 1000 steps. 

During inference, initial samples are drawn $X_0[j] \sim \mathcal{N}(0, 1) \textbf{ if } M[j]=0$ where $M$ corresponds to a binary conditioning mask (described below). All other $X_0[j]$ are set to their true (unmasked) values determined by simulation or experiment. A neural network model (described below) iteratively predicts how to to remove noise from the sample, given the mask and diffusion time $\hat{\epsilon} = f_\theta(X_t, M, t)$. The sample is updated according to $X_{t+1} = \frac{1}{\sqrt{\alpha}} \left(X_{t} - \frac{1 - \alpha}{\sqrt{1 - \bar{\alpha}}}\hat{\epsilon} \right) + \sqrt{(1-\alpha)}\epsilon$ and noise is continuously removed over 1000 steps. Both image and scalar quantities are treated with identical (de)noising procedures, the only difference being the dimensions of data modalities. 

\subsubsection*{Neural Network Architecture}
A multi-modal neural network architecture is used to encode and predict noise in both scalar and image pixel space. We adapt a U-net architecture~\cite{ronneberger2015u} developed by~\cite{dome272_diffusion_models_pytorch} to ingest and predict multi-modal data. Each image view is embedded by independent convolutions networks containing 48 channels and 3 up/down sampling layers. Both input scalars, output scalars, and mask tokens are embedded by fully connected neural networks containing one 64-dimensional hidden layers. Diffusion time is embedded by a sin/cos positional encoding. Scalar, mask, and time embeddings are concatenated and added to U-net embedding at each up and down sampling step. Image embedding for each view receive the same time, scalar, and mask conditioning but are independent from each other until the bottleneck layer (lowest dimension of the U-net). At the bottleneck layer outputs from each image view are pooled along the channel dimension, flattened, and concatenated along with scalar and mask embedding. Output and input scalar decoders take the flattened bottleneck layer as input and predict noise for each scalar. Image decoders pool information from the bottleneck layer and upsample three times to recover the original pixel and channel dimensions.

Models are constructed in PyTorch~\cite{paszke2019pytorch}, and trained using the Adam optimizer~\cite{KingmaB14}. We use a batch size of 256 and learning rate of 0.0003. An exponential moving average with constant 0.995 is applied to weights. Additional architecture details and hyper-parameters are included in Tables S1 and S2).

\subsubsection*{Masking}
Binary masks inform the model if a conditioning variable (scalar or image) is provided for conditioning, or if it should be predicted during the denoising (inference) process. Certain masks are linked to others if those quantities can never be known without the other. Specifically, all inputs masks are linked since they must all be specified to perform a simulation and cannot be partially observed in experiment. The four inferred outputs scalars ($\rho$R Weighted Harmonic Mean, $\rho$R Mean, RKE at Bang-time, and RKE at Minimum Volume) are linked for the same reason~\cite{hurricane2022extensions}. The three velocity scalars are linked because they are x,y,z components of the same vector. All other output scalars have individual masks along with each image view and channel. During training, all independent masks are sampled according to a Bernoulli distribution with constant 0.5. Ablations were performed with varied constant values and with curriculum learning strategies to gradually lower the constant as function of training time, but a fixed constant of 0.5 provided optimal performance across forward, inverse, and imputation tasks.







\begin{figure}
\centering
\includegraphics[width=1.0\linewidth]{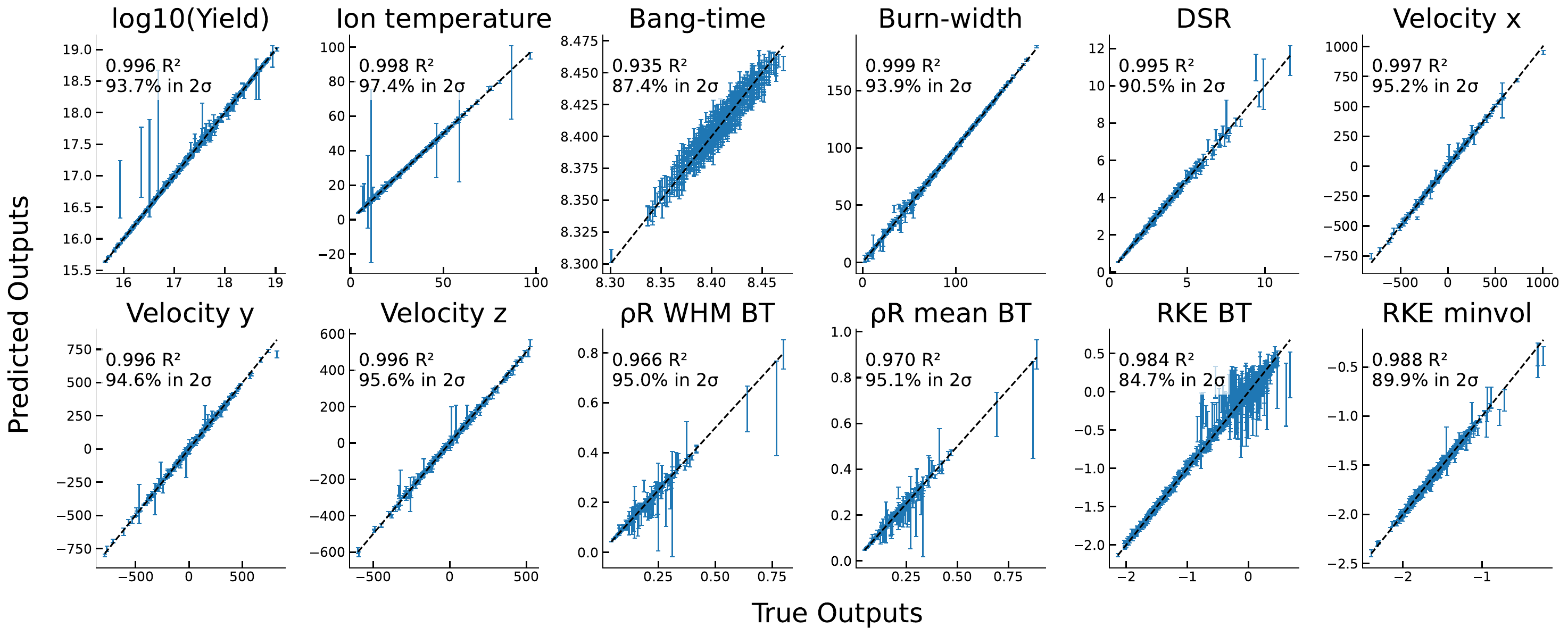}
\caption{Predicting scalar outputs from simulations inputs. For each test simulation, 10 predictions are made. Error bars show the mean and standard deviation across these 10 predictions. $R$2 metrics are computed between the predicted means and the simulated ground truth. The percent of ground truth samples within two standard deviations of the predicted distribution is also shown for each scalar.}
\label{fig:outputs_all}
\end{figure}

\begin{figure}
\centering
\includegraphics[width=1.0\linewidth]{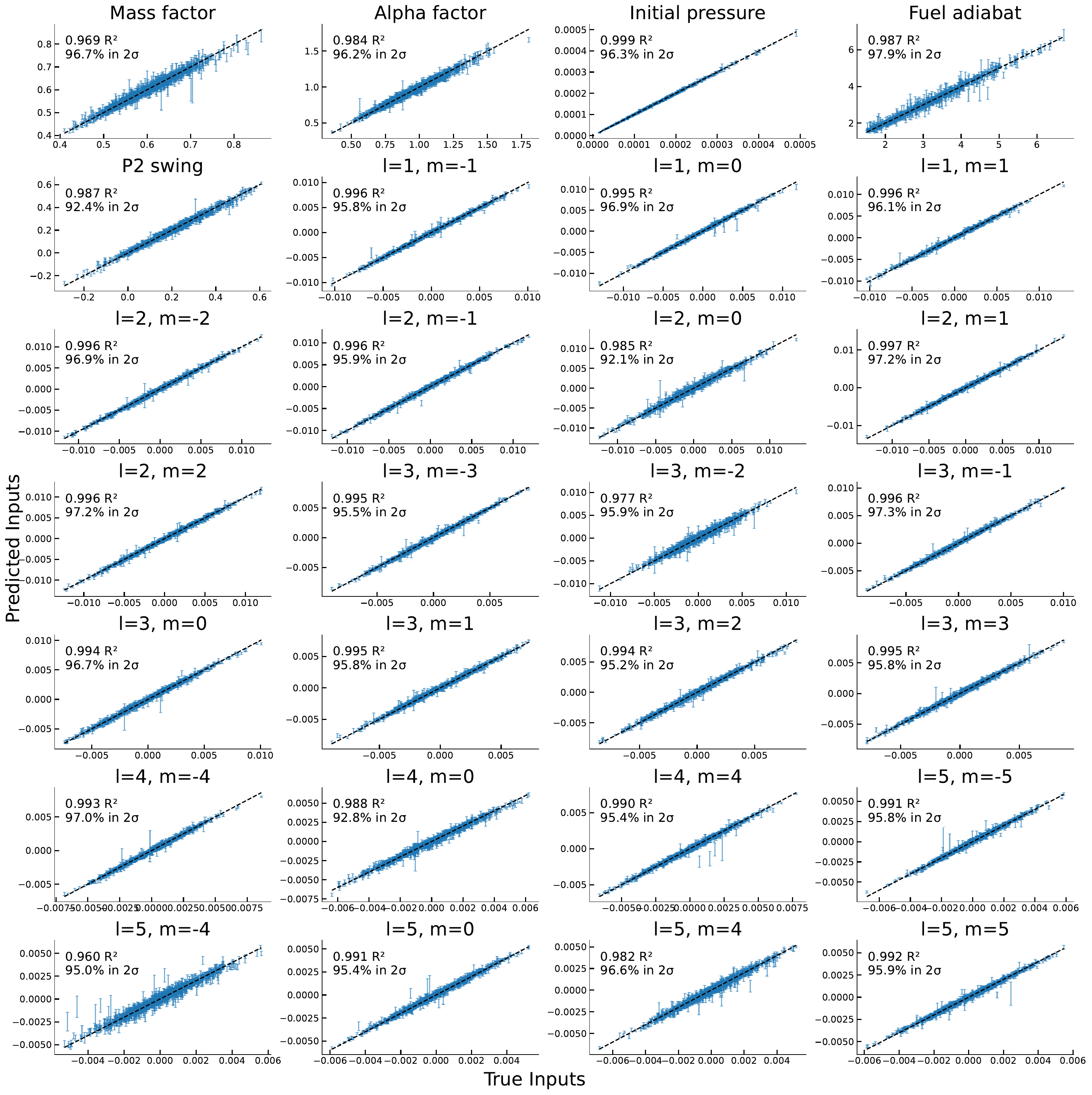}
\caption{Predicting scalar inputs from simulations outputs. For each test simulation, 10 predictions are made. Error bars show the mean and standard deviation across these 10 predictions. $R$2 metrics are computed between the predicted means and the simulated ground truth. The percent of ground truth samples within two standard deviations of the predicted distribution is also shown for each scalar.}
\label{fig:inputs_all}
\end{figure}
\newpage

\begin{figure}
\centering
\includegraphics[width=1.0\linewidth]{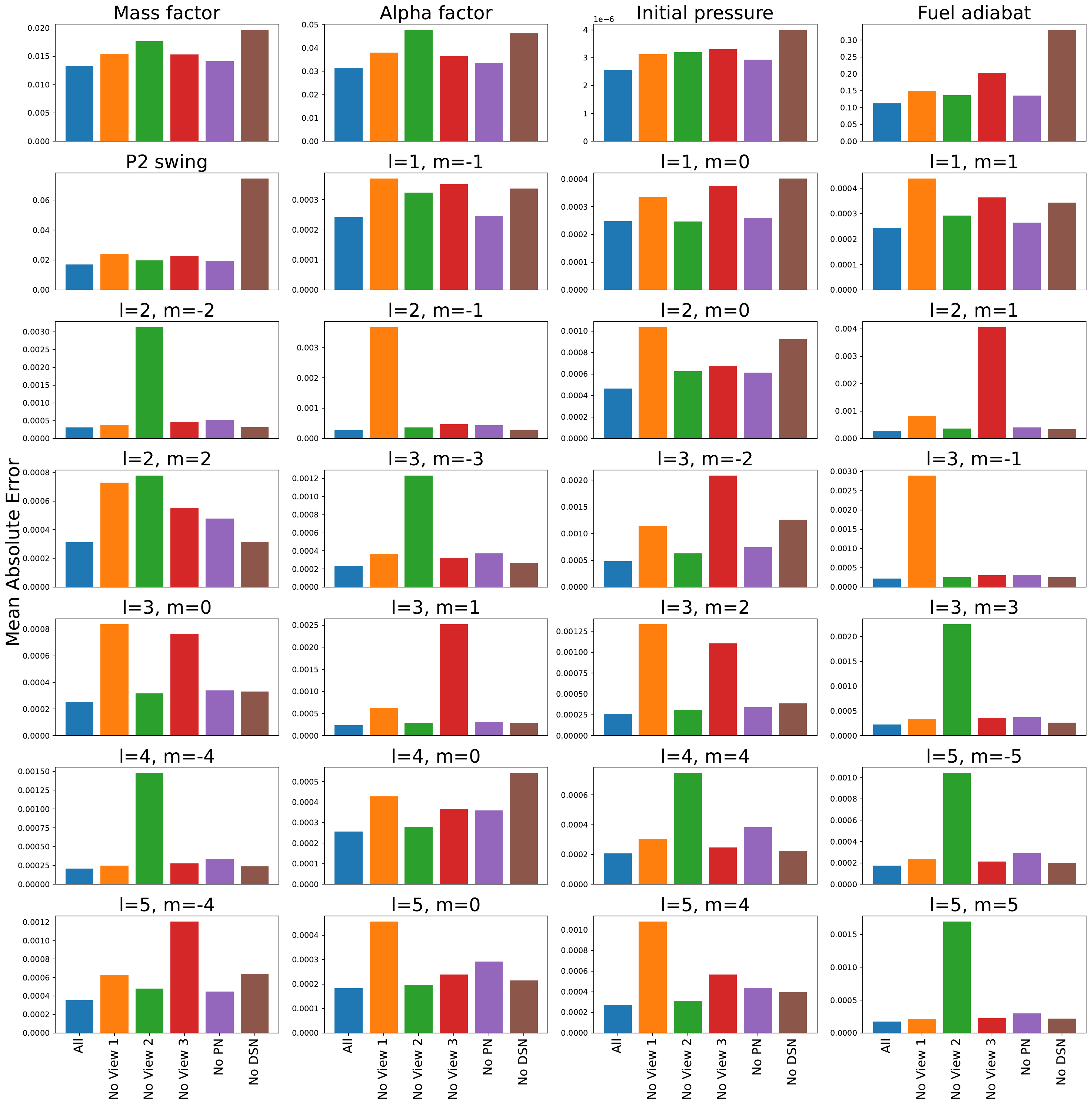}
\caption{Mean average error over each input when model is provided all outputs (blue) and when removing partial image information (each view and each channel)}
\label{fig:mae_all}
\end{figure}

\begin{figure}[h]
\centering
\includegraphics[width=0.9\linewidth]{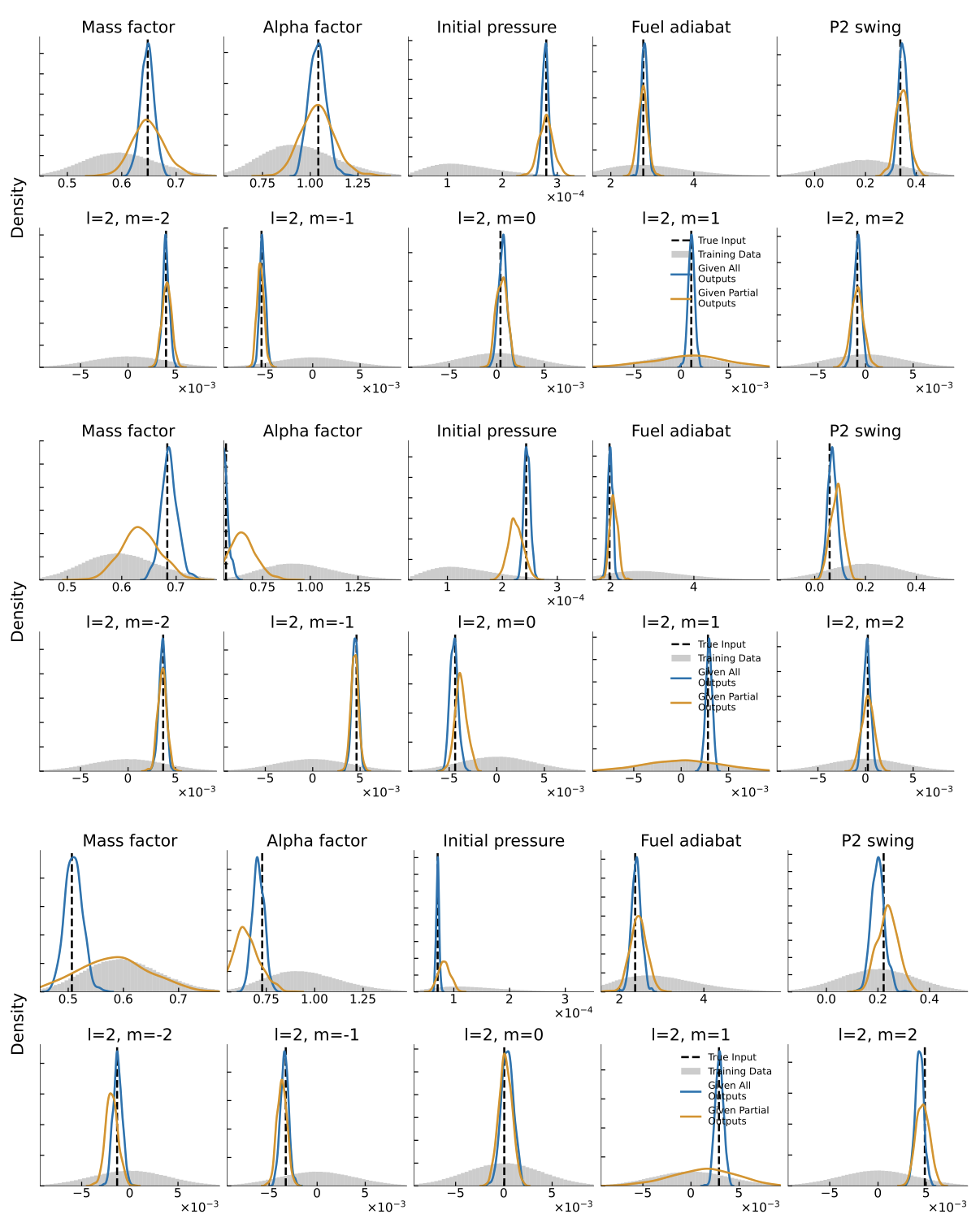}
\caption{Input distributions for three additional test samples. Distribution of inputs for 500 generations when providing model all outputs (blue) or half of outputs (orange). Ground truth inputs are shown by black dashed line. Distribution of all training data is shown in gray.}
\label{fig:fig5_inps_combined}
\end{figure}
\newpage

\begin{figure}[h]
\centering
\includegraphics[width=0.9\linewidth]{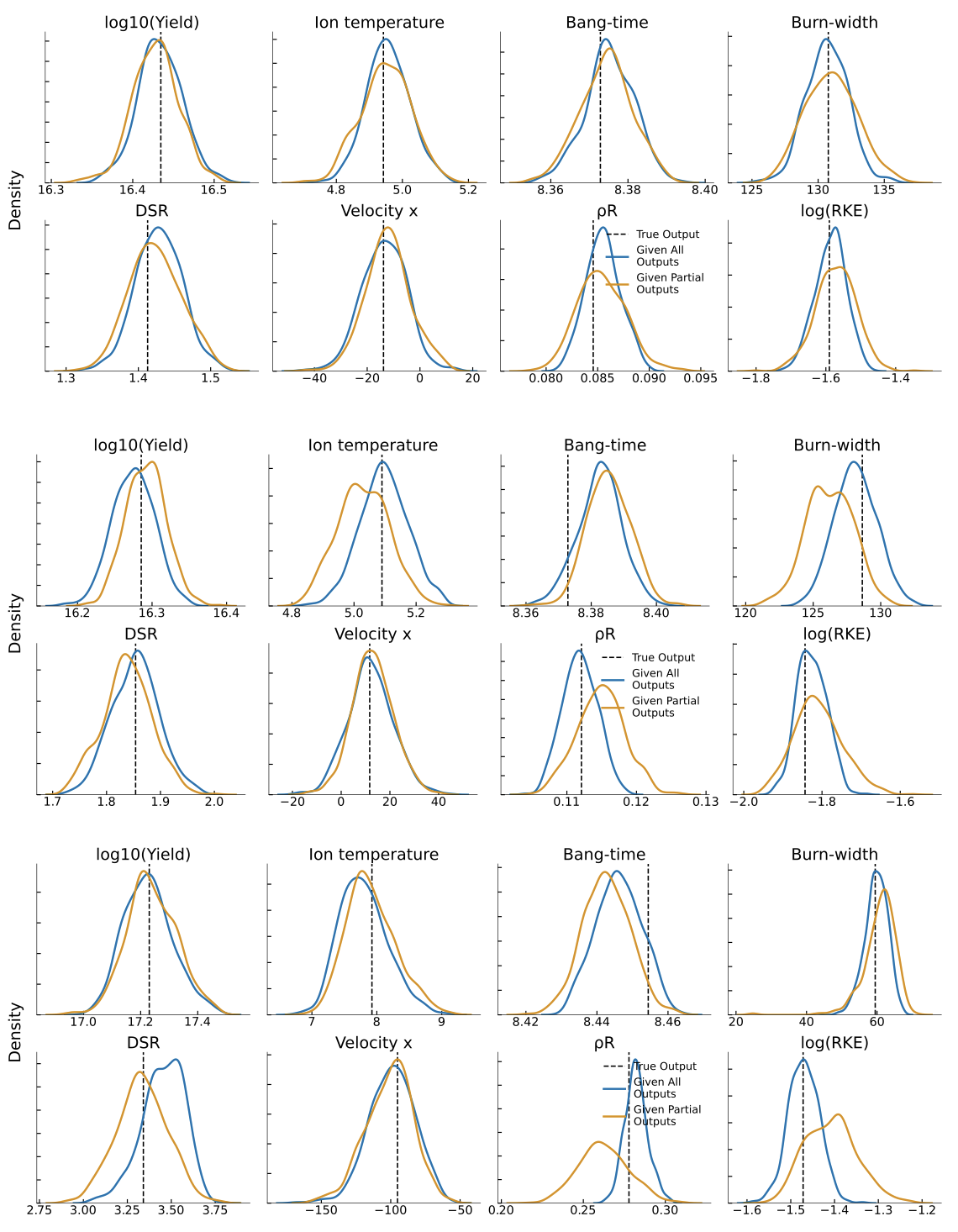}
\caption{Round-trip output distributions for three additional test samples. Distributions are produced by running predicted input distributions through the forward model. Original conditioning contained all outputs (blue) or half of outputs (orange).}
\label{fig5_outs_combined}
\end{figure}
\newpage


\clearpage


\begin{table}[h]
\centering
\begin{tabular}{l l c}
\hline
\textbf{Name} & \textbf{Units} & \textbf{Size} \\
\hline
Initial pressure & kbar & 1 \\
Alpha heating multiplier & -- & 1 \\
Shocked mass fraction & -- & 1 \\
Non-radial flow term & -- & 1 \\
Dense fuel adiabat & -- & 1 \\
$\langle l, m \rangle$ symmetry modes & -- & 23 \\
\hline
\end{tabular}
\caption{\textbf{Rocket-Piston input quantities, dimensions, and units.}}
\end{table}

\begin{table}[h]
\centering
\begin{tabular}{l l c}
\hline
\textbf{Name} & \textbf{Units} & \textbf{Size} \\
\hline
Log(Yield) & -- & 1 \\
Ion Temperature & keV & 1 \\
Bang-Time & ns & 1 \\
Burn-Width & $\mu$m & 1 \\
Down-scatter Ratio & -- & 1 \\
Hot-spot Velocity Vector & km/s & 3 \\
$\rho$R Weighted Harmonic Mean & g/cm$^2$ & 1 \\
$\rho$R Mean & g/cm$^2$ & 1 \\
Residual Kinetic Energy at Minimum Volume & kJ & 1 \\
Residual Kinetic Energy at Bang-Time & kJ & 1 \\
Primary Image & -- & 3x32x32 \\
Down-scattered Image & -- & 3x32x32 \\
\hline
\end{tabular}
\caption{\textbf{Rocket-Piston output quantities, dimensions, and units.}}
\end{table}

\begin{table}
\centering
\caption{\textbf{Input and output scalar diffusion with fixed forward and inverse models} We evaluate how task-specific models compare to the unified JointDiff model, considering both forward and inverse modeling tasks with all inputs and outputs specified. As baselines, we use simplified versions of JointDiff: one that makes direct predictions without the diffusion objective, one that uses diffusion for the forward task only, and one that uses diffusion for the inverse test only. All baseline models are trained for the same number of epochs and with the same hyperparameters as the combined model. We report $R^2$ values for scalar predictions and the percentage of true values captured within the predicted standard deviation. All results generated over 986 test simulations with 10 diffusion samples each. The first model is deterministic and therefore only one prediction is made.}
\label{tab:baselines}
\begin{tabular}{lrrrr}
Model & Output $R^{2}$ & Input $R^{2}$ & Output 2$\sigma$ & Input 2$\sigma$ \\
\hline
1. Fwd+Inv & 0.998 & 0.995 & --   & -- \\
2. Fwd+Diffusion     & 0.998 & --    & 97.6 & -- \\
3. Inv+Diffusion     & --    & 0.997 & --   & 96.4 \\
4. Fwd+Inv+Diffusion  & 0.998 & 0.992 & 97.4 & 96.7 \\
\hline
\end{tabular}
\end{table}





\clearpage 





\end{document}